\documentclass[prl,twocolumn,showpacs,preprintnumbers,amsmath,amssymb,superscriptaddress]{revtex4}

\usepackage{graphicx}
\usepackage{natbib}
\usepackage{latexsym}
\begin{document}

\newcommand{\ei}{\hat{a}}
\newcommand{\eidag}{\hat{a}^{\dag}}
\newcommand{\hn}{\hat{n}}

\newcommand{\ddt}{\frac{d}{dt}}
\newcommand{\Lx}{\hat{L}_x}
\newcommand{\Ly}{\hat{L}_y}
\newcommand{\Lz}{\hat{L}_z}
\newcommand{\Li}{\hat{L}_i}
\newcommand{\Lj}{\hat{L}_j}
\newcommand{\Lk}{\hat{L}_k}
\newcommand{\Lp}{\hat{L}_+}
\newcommand{\Lm}{\hat{L}_-}
\newcommand{\Lpm}{\hat{L}_\pm}

\newcommand{\Dxx}{\Delta_{xx}}
\newcommand{\Dyy}{\Delta_{yy}}
\newcommand{\Dzz}{\Delta_{zz}}
\newcommand{\Dxy}{\Delta_{xy}}
\newcommand{\Dyz}{\Delta_{yz}}
\newcommand{\Dzx}{\Delta_{zx}}

\title{Bosonic amplification of  noise-induced suppression of phase diffusion}\author{Y. Khodorkovsky}
\affiliation{Department of Chemistry, Ben-Gurion University of
the Negev, P.O.B. 653, Beer-Sheva 84105, Israel}
\author{G. Kurizki}
\affiliation{Department of Chemical Physics, The Weizmann Institute of Science, Rehovot 76100, Israel}
\author{A. Vardi}
\affiliation{Department of Chemistry, Ben-Gurion University of
the Negev, P.O.B. 653, Beer-Sheva 84105, Israel}

\begin{abstract}
We study the effect of noise-induced dephasing on collisional phase-diffusion in the two-site Bose-Hubbard model. Dephasing of the quasi-momentum modes may slow down phase-diffusion in the quantum Zeno limit. Remarkably, the degree of suppression is enhanced by a bosonic factor of order $N/\log{N}$ as the particle number $N$ increases. 
\end{abstract}  
\maketitle

The interplay between unitary evolution and decoherence has been a central issue of quantum mechanics. A universal formula has been put forward for the dynamical control of quantum systems weakly coupled to a bath \cite{Kofman}. According to this formula, relaxation or decoherence can either be suppressed or enhanced by interventions whose rate is much higher than, or comparable to (respectively) the inverse non-Markovian memory time of the bath response. If these interventions are either projective measurements or, equivalently, stochastic dephasing of the system's evolution \cite{Kofman}, the resulting slow-down or speed-up of the relaxation/decoherence coincide with the quantum Zeno effect (QZE) \cite{Kofman,QZE} or the anti-Zeno effect (AZE) \cite{Kofman,AZE}, respectively.

This universal formula provides simple recipes for the dynamical projection of both single- and multi-partite quantum states, provided the bath spectral response is known. Yet an open question remains: can we similarly control/protect quantum states of large multi-partite systems from the buildup of {\sl many body correlations} among its $N$ interacting particles ? The analysis of this scenario is tantalizing and nearly impossible for a multi-mode system with large $N$. However, useful insights can be gained by exploring few-mode models, for which full numerical solutions may be used to support analytic early-time approximations. At such times, we may describe the slowdown or speedup of the system's many-body evolution by noisy perturbations, as QZE or AZE, respectively. Since phase-diffusion between atomic Bose-Einstein condensates (BEC) \cite{PhaseDiffusion,Vardi01,Greiner02,ChipInterference} has a non-Markovian correlation time of ms, it is particularly amenable to the observation of the QZE and AZE \cite{MazetsStreed}. Our main result here, is that for $N$-boson condensates, the QZE is {\sl Bose amplified}.

\begin{figure}
\centering
\includegraphics[width=0.5\textwidth]{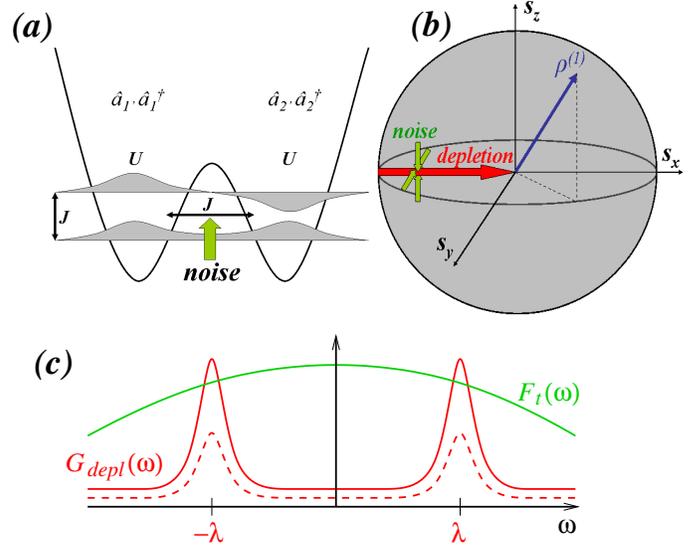}
\caption{(color online) Bose-enhanced QZE suppression of two-mode depletion: (a) Schematic diagram of the two-site Bose-Hubbard model. Depletion of the excited quasimomentum state is slowed down by a non-local noise source. (b) $N$-partite Bloch sphere, with $\Lz$-dependent depletion and $\Lx$-dependent noise. (c) Suppression of the depletion rate $R$ as the overlap of $G_{depl}(\omega)$ and $F_t(\omega)$ noise decreases with $N$ (solid red to dashed red - Eqs. (\ref{Gershonstuff}),(\ref{g12Gershon})).}
\label{qzeprl1_fig1}
\end{figure}

Specifically, we consider the noise-induced suppression of phase-diffusion in the two-site Bose-Hubbard model, recently used to describe experiments of quantum interference \cite{ChipInterference,JoChoi07} and tunneling in an array of double wells \cite{doublewells}. This model, under the tight-binding condition for 3D wells, $l \gg N|a|$, where $a$ is the scattering length, and $l=\sqrt{\hbar/(m\omega_0)}$ is the characteristic size of a trap with frequency $\omega_0$, is accurately described by the quantized two-mode Josephson Hamiltonian \cite{Makhlin01}, here rewritten as
\begin{equation}
\label{Ham}
\hat{H}=-J\Lx+U\Lz^2~,
\end{equation}
where $\Lx=(\eidag_1 \ei_2+\eidag_2\ei_1)/2$, $\Ly=(\eidag_1\ei_2-\eidag_2\ei_1)/(2i)$, and $\Lz=(\hn_1 - \hn_2)/2$ generate the $SU(2)$ Lie algebra. The mean-field values $\langle\Li\rangle$ determine the reduced single-particle density matrix $\rho^{(1)}=\langle \eidag_i \ei_j \rangle/N$, with mode indices $i,j=1,2$. The operators $\ei_i$ and $\eidag_i$ are bosonic annihilation and creation operators respectively, for particles in mode $i$ with corresponding particle number operators $\hn_{i}=\eidag_{i}\ei_{i}$. The bias potential is here set to zero, $J$ is the intermode coupling, and $U$ is the collisional interaction frequency. We have eliminated $c$-number terms, proportional to the conserved total particle number $N=\hn_1+\hn_2$.   

The collisional $\Lz^2$ term in the Hamiltonian (\ref{Ham}) leads to 'phase-diffusion' \cite{PhaseDiffusion,Vardi01,Greiner02,ChipInterference} which degrades the reduced single-particle coherence. Its eigenstates,
\begin{equation}
\left|l,m\right\rangle=\frac{1}{\sqrt{\left(l+m\right)!\left(l-m\right)!}}\left(a_1^\dag\right)^{l+m}\left(a_2^\dag\right)^{l-m}|0\rangle,
\label{fock}
\end{equation} 
constitute a preferred basis set resilient to this process \cite{Zurek03}. On the other hand, the most sensitive states to phase-diffusion are the spin coherent states 
$
|\theta,\phi\rangle=e^{-i\phi\Lz}e^{-i\theta\Ly} |l,-l\rangle\nonumber,
$ 
with $\theta=\pi/2$, corresponding to equal populations of the two sites, and a well-defined relative phase $\phi$. For fully separated modes ($J=0$) the single particle coherence of these states is lost as $\exp{\left[-(t/t_d)^2\right]}$ with a characteristic decay time $t_d=(U\sqrt{l})^{-1}$ and revives after $t_r=\pi/U$ \cite{PhaseDiffusion,Greiner02}. For finite $J$ and $U>0$ (repulsive interaction) the fastest phase-diffusion occurs for the antisymmetric coherent state
\begin{equation}
\left|\pi/2,\pi\right\rangle=\frac{1}{2^l}\sum_{m=-l}^l (-1)^{l+m}
\left(
\begin{array}{c}
2l\\
l+m
\end{array}\right)^{1/2}|l,m\rangle~,
\label{oddstate}
\end{equation}
i.e. the state with all particles populating the excited, odd superposition of the modes. This experimentally realizable state \cite{ChipInterference} will be used as the initial condition throughout this work. Its evolution with $U>0$ is identical to the evolution of the state $|\pi/2, 0\rangle$ with $U<0$, which for $|UN|>J$ drives the system towards a macroscopic cat state \cite{Cats}. Since $[\hat{{\bf L}}^2,\hat{H}]=0$, we fix $l=N/2$.   

\begin{figure}
\centering
\includegraphics[width=0.48\textwidth]{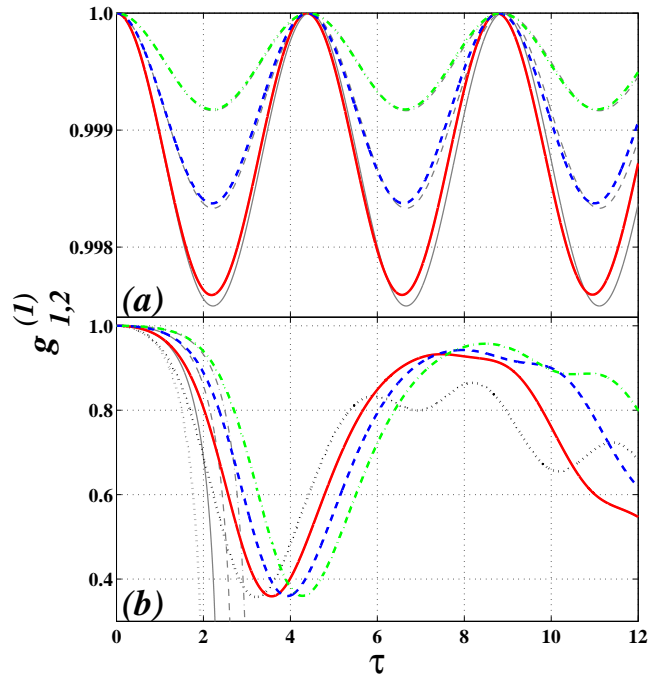}
\caption{(color online) Single-particle coherence as a function of rescaled time, starting from the coherent state $|\pi/2,\pi\rangle$ with (a) $\kappa=0.5$, (b) $\kappa=2$. Bold lines in (a) correspond to $N=100$ (solid red), $150$ (dashed blue), and $300$ (dash-dotted green) particles. Bold lines in (b) are $N=50$ (dotted black), $100$ (solid red), $200$ (dashed blue), and $400$ (dash-dotted green) particles. Gray lines correspond to the analytic forms of Eq.~(\ref{gdyn}) for the same $N$.} 
\label{qzeprl1_fig2}
\end{figure} 

Control of the phase diffusion of the state (\ref{oddstate}) may be attained by noise-induced dephasing of the odd- and even two-mode superpositions, as illustrated in Fig.~\ref{qzeprl1_fig1}. This control can be introduced by any noise source that does not distinguish between the sites, e.g. an off-resonant incoherent light source focused at the barrier between them (Fig.~\ref{qzeprl1_fig1}a). Such $\Lx$ noise, implemented perpendicular to the depletion axis $\Lz$ (Fig.~\ref{qzeprl1_fig1}b), affects the stochastic (noisy) modulation of the splitting $J$, modifying the system Hamiltonian as
\begin{equation}
\hat{H}_S(t)=[J+\hbar \delta_S(t)]\Lx+U\Lz^2
\label{modulation}
\end{equation} 
The depletion-dephasing interplay is described by the general non-Markov (Zwanzig-Nakajima) second-order master equation (ME) for the reduced density matrix of the system. The resulting noise-controlled, normalized decoherence rate $R$ conforms to the universal formula \cite{Kofman},
\begin{eqnarray}
\label{Gershonstuff}
R(t)&=&2\pi \int_{-\infty}^{\infty} F_t(\omega)G_{depl}(\omega+J/\hbar)d\omega~,\\
F_t(\omega)&=&(2/N)^2\langle |\epsilon_t(\omega)|^2 \Lx^2\rangle,~\nonumber\\
G_{depl}(\omega+J/\hbar)&=&\frac{UN}{4\pi}\int_{-\infty}^{\infty}\Phi_{depl}(t) e^{i(\omega+J/\hbar)t} dt,\nonumber
\end{eqnarray}
expressing $R(t)$ as the convolution of the finite time spectral intensity of the noise control (where $\epsilon_t(\omega)=\int_0^t dt' e^{i[\omega t' + \int_0^{t'}\delta_s(t'',\ldots)dt'']}$ is the Fourier transform of the stochastic phase factor), and the Fourier transform of the depletion correlation function $\Phi_{depl}=(2/N)^2\langle e^{i J\Lx(t-t')/\hbar}\Lz(t) e^{-i J\Lx(t-t')/\hbar}\Lz(t')\rangle$. Equation (\ref{Gershonstuff}) provides a general recipe for controlling quantum depletion by noise. It yields the QZE limit of $R(t)$ suppression when $F_t(\omega)$ is spectrally much broader than $G_{depl}$ (Fig.~\ref{qzeprl1_fig1}c), i.e. when the inverse width of these spectral functions (their memory times) satisfy $(t_c)_{noise}\ll (t_c)_{depl}$. Conversely, it yields the AZE limit of $R(t)$ enhancement when the two spectral widths or memory times are comparable and their spectral centers are mutually shifted. 

In what follows, we focus on the QZE limit, neglecting $(t_c)_{noise}$ altogether, i.e. taking the broadband noise to be Markovian. In atomic BECs, this limit is obtained for $1/(t_c)_{noise}\gg \Gamma_x \gg 1/(t_c)_{depl} \gtrsim kHz$, $\Gamma_x$ being the rate of the Markovian dephasing. In this limit, we can use the Markovian quantum kinetic ME
\begin{equation}
\dot{\hat{\rho}}=i\left[\hat{\rho},\hat{H}\right]-\Gamma_x\left[\Lx,\left[\Lx,\hat{\rho}\right]\right]~.
\label{master}
\end{equation} 

Exact solutions of Eq.~(\ref{master}) can be found by expansion in the $|l,m\rangle$ basis set and numerical integration. In order to analytically approximate the initial phase-diffusion, we truncate the hierarchy of dynamical equations for the $\Li$ operators, at second-order correlations to obtain the Bogoliubov Backreaction (BBR) equations \cite{Vardi01,Tikhonenkov07}, for the mean-field single-particle Bloch vector ${\bf s}=2\langle\hat{\bf L}\rangle/N$ and the correlation functions $\Delta_{ij}=4(\langle\Li\Lj+\Lj\Li\rangle-2\langle\Li\rangle\langle\Lj\rangle)/N^2$. Linearizing these equations around the $|\pi/2,\pi\rangle$ state, we obtain the initial dynamics of the normalized correlation function $g^{(1)}_{1,2}=\left|\rho^{(1)}_{12}\right |\left(\rho^{(1)}_{11}\rho^{(1)}_{22}\right)^{-1/2}$, corresponding to the fringe visibility in interference experiments \cite{ChipInterference}. In the absence of noise ($\Gamma_x=0$) we find, 
\begin{equation}
\left({g^{(1)}_{1,2}}(\tau)\right)^2=\left\{
\begin{array}{lr}
1-\frac{4\cot^2(2\Theta)}{N}\sin^2(\lambda\tau) &\kappa<1\\
~&~\\
1-\frac{4\coth^2(2i\Theta)}{N}\sinh^2(\lambda\tau) &\kappa>1
\end{array}\right. ,
\label{gdyn}
\end{equation}
where $\lambda=\left |\sqrt{1-\kappa}\right |$ and $\tan\Theta=\sqrt{1-\kappa}$, $\kappa=UN/J$ is the coupling parameter, and $\tau=Jt$ is the rescaled time. Dynamical BEC depletion in the weak-interaction ($\kappa<1$) regime is thus bound and inversely proportional to the number of particles $N$. By contrast, for strong interactions ($\kappa>1$), the phase-diffusion rate is independent of the number of particles, but its onset time scales logarithmically with $N$ \cite{Vardi01}. This strong interaction instability may account for the rapid heating observed in the merging of two condensates with a $\pi$ relative-phase, on an atom chip \cite{JoChoi07}.

\begin{figure}
\centering
\includegraphics[width=0.5\textwidth] {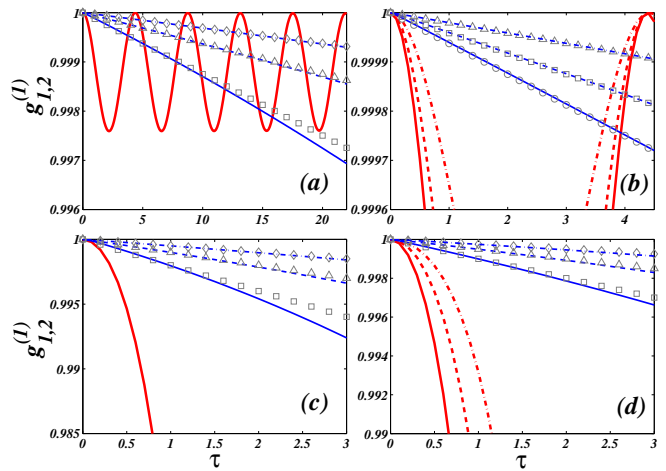}
\caption{(color online) Single-particle coherence as a function of rescaled time, with $\kappa=0.5$ in (a),(b) and $\kappa=2$ in (c),(d). The Hamiltonian dynamics with $\gamma_x=0$ (bold red lines) is compared to the evolution with non-vanishing $\gamma_x$ (normal blue lines). Solid, dashed, and dash-dotted blue lines in (a) and (c) correspond respectively to $\gamma_x=5,10,20$ and $N=100$. Normal blue lines in (b) and (d) correspond to $\gamma_x=10$. Solid, dashed and dash-dotted lines in (b) correspond to $N=100,150,300$ particles, respectively, whereas solid, dashed and dash-dotted lines in (d) are $N=100,200,400$ particles, respectively. Gray symbols portray the QZE behavior of Eq.~(\ref{gcontzeno}).}
\label{qzeprl1_fig3}
\end{figure}  

Numerical results based on the Markovian ME confirm the short-time dynamics of Eq.~(\ref{gdyn}). The weak-coupling behavior (Fig.~\ref{qzeprl1_fig2}a) exhibits the anticipated stable oscillations. The oscillation amplitude decreases with increasing $N$ while keeping $\kappa$ fixed, so that ${\cal O}(1/N)$ initial quantum fluctuations remain small compared to the ${\cal O}(1)$ classical mean-field. By contrast, for strong interactions (Fig.~\ref{qzeprl1_fig2}b), quantum fluctuations grow rapidly and single-particle coherence is lost. Equation (\ref{gdyn}) gives an accurate description of the depletion at short times and a good estimate for the phase-diffusion time. At longer times, depletion is significant and the linearized BBR equations are no longer adequate. The fully nonlinear BBR equations, however, describe the loss of single-particle coherence with good accuracy. Partial and full revivals are observed due to the finite number ($N+1$) of phase-space dimensions.

We now proceed to explore the effect of noise on phase diffusion. It is evident from Eq.\ (\ref{gdyn}) that in the absence of noise, regardless of the interaction strength, $g^{(1)}_{1,2}$ at $\tau< 1/\lambda$ scales {\it quadratically } rather than linearly in time,
\begin{equation}
g^{(1)}_{1,2}(\tau)=\frac{2\left\langle\Lx\right\rangle}{N}=1-(\Omega\tau)^2,
\end{equation}
where $\Omega=\sqrt{2/N}|\coth(2i\Theta)|\lambda$. When the frequent measurement dephasing condition $\gamma_x\gg\lambda>\Omega$ is satisfied, we can adiabatically eliminate $\Dyz$ in the linearized BBR equations and obtain the QZE behavior, 
\begin{equation}
g^{(1)}_{1,2}(\tau)=\exp\left(-\frac{\Omega^2}{2\gamma_x}\tau\right), 
\label{gcontzeno}
\end{equation}
 where $\gamma_x=\Gamma_x/J$. The modified diffusion time in the presence of noise, $\tilde{\tau}_{d}=2\gamma_x / \Omega^2=N\gamma_x\tanh^2(2i\Theta) / \lambda^2\gg 1 / \Omega$, should be compared, in the strong-interaction regime, with the noise-free, finite-$J$ diffusion time, $\tau_{d}\propto\log\left[N\tanh^2(2i\Theta)\right] / \lambda$. Hence, the transition from the hyperbolic growth (\ref{gdyn}) which only depends on $N$ through its onset time, to the QZE-suppressed depletion (\ref{gcontzeno}) at a rate linear in $N$, introduces a bosonic factor of order $N/\log{N} \gg 1$ in the diffusion time ratio $\tilde{\tau}_{d}/\tau_{d}$. The QZE is thus strongly amplified with increasing particle number. This constitutes the main result of this work.

Similar scaling is obtained from the universal Eq.~(\ref{Gershonstuff}). In the QZE limit of non-Markovian depletion in the presence of Markovian noise, $g_{1,2}^{(1)}(t)\simeq 1-\int_{0}^{t}R(t')dt'$ may be inferred from it by substitution of the stochastic broadband $F_t(\omega)\simeq(2/N)^2(t/2\pi)\mbox{sinc}^2(\omega t/2)\langle\Lx^2\rangle$, so that
\begin{equation}
R(t)\simeq \frac{2t}{\pi N^2}\int_{-\infty}^{\infty}G_{depl}(\omega +J/\hbar)\mbox{ sinc}^2\left(\frac{\omega t}{2}\right)\langle\Lx^2\rangle d\omega~,
\label{g12Gershon}
\end{equation}
This form characterizes frequent projective measurements \cite{AZE}. The QZE is obtained when $\mbox{ sinc}^2(\omega t/2)\langle \Lx^2\rangle$ is much broader than $G_{depl}$. As $N$ is increased while keeping $\kappa$ fixed,  we have $4\langle \Lx^2\rangle/N^2\approx s_x^2\sim 1$ and $G_{depl}(\omega +J/\hbar)\propto 1/N$ (Fig.~\ref{qzeprl1_fig1}c) so that $R(t)\propto 1/N$.

The QZE suppressed phase diffusion is illustrated in the numerical results of Fig.~\ref{qzeprl1_fig3}, where we compare the initial evolution of $g^{(1)}_{1,2}$ with and without noise, in the weak- and strong-interaction regimes. The weak-interaction oscillations of Eq.~(\ref{gdyn}) are replaced, as $\gamma_x$ is increased, by the exponential decay of Eq.~(\ref{gcontzeno}), at a rate proportional to $1/(N\gamma_x)$ (Figs.~\ref{qzeprl1_fig3}a,b). The strong-interaction dependence on the dephasing rate $\gamma_x$ (Fig.~\ref{qzeprl1_fig3}c) and its Bose-amplified suppression (Fig. \ref{qzeprl1_fig3}d), show a clear transition from $\log N$ dependent diffusion-times followed by $N$-independent depletion rate, to $1/N$ dependent depletion rates. These numerical results agree well with the appropriate closed form of Eq.~(\ref{gdyn}) and Eq.~(\ref{gcontzeno}).

\begin{figure}
\centering
\includegraphics[width=0.5\textwidth] {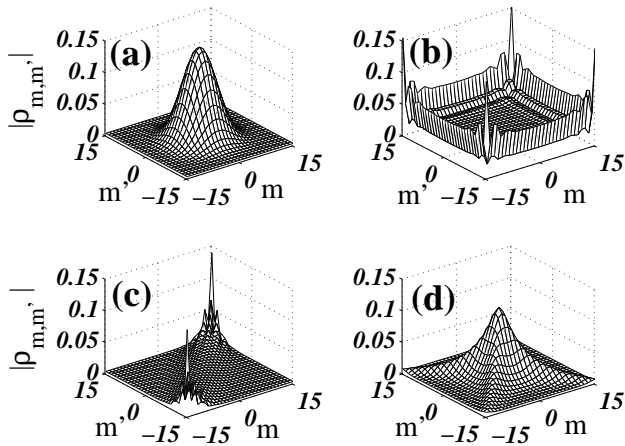}
\caption{(color online) Absolute values of the elements of $\rho$ in relative Fock basis $|l,m\rangle$, for $N=30$ and $\kappa=2$: (a) Initial coherent state, (b) cat state after noise-free evolution at $\tau=2.4$, (c) incoherent mixture after evolution with local site noise with $\Gamma_z=0.05J$, at $\tau=2.3$, (d) protected single-particle coherence after evolution with $\Gamma_x=J$.} 
\label{qzeprl1_fig4}
\end{figure}  

It is interesting to contrast this noise-induced QZE behavior to the effects of local-site noise, which may be induced by collisions with thermal particles \cite{Anglin97}. In Fig.~\ref{qzeprl1_fig4} we plot the density matrix elements for the intial coherent state (Fig.~\ref{qzeprl1_fig4}a) and for the macroscopic cat state generated after noise-free evolution in the strong interaction regime (Fig.~\ref{qzeprl1_fig4}b). The dynamics leading to this cat state is extremely sensitive to local site noise  \cite{Cats}, which destroys the macroscopic coherence, resulting in a 50-50 statistical mixture (Fig.~\ref{qzeprl1_fig4}c). Phase-diffusion in this case, is {\it enhanced} to the extent that the strong-interaction diffusion-time is bound at large $N$ \cite{Vardi01}. By contrast, the site-indiscriminate (non-local) noise considered here protects the single-particle coherence and slows down phase-diffusion (Fig.~\ref{qzeprl1_fig4}d). Whereas any weak local noise will degrade the intricate dynamics leading to a cat state, our non-local noise needs to be sufficiently strong to induce the QZE.    

To conclude, we have found novel collective features of the QZE, which do not appear in the noise-controlled decay of single particles. The results presented here are generic rather than specific to the two-site Bose-Hubbard model of atomic BECs.  There is currently  great interest in phase-diffusion experiments, enabling the measurement of single-particle coherence via the visibility of interference fringes \cite{Greiner02,ChipInterference,JoChoi07}. Our predictions may  thus be directly verified using current experimental apparata. Consequently, new avenues may be opened for noise-control of complex multipartite systems. 
  
This work was supported by the Israel Science Foundation (Grant Nos. 8006/03, 582/07) and the Minerva Foundation through a grant for a junior research group. G. K. is supported by GIF, DIP and EC (MIDAS, STREP).

\end{document}